\documentstyle[12pt]{article}
\begin{document}

\centerline{\Large {\bf Heisenberg $XYZ$ Hamiltonian 
with Integrable Impurities}} 
\bigskip
\centerline{\bf Zhan-Ning Hu\footnote{\bf E-mail: 
huzn@aphy.iphy.ac.cn}}
\smallskip
\centerline{Institute of Physics and Center 
for Condensed Matter Physics,} 
\centerline{Chinese Academy of Sciences, Beijing 100080, China}

\begin{center}
\begin{minipage}{5in}
\centerline{\large\bf   Abstract}

In this letter, a Hamiltonian of the impurity model is constructed 
within 
the framework of the open boundary Heisenberg $XYZ$ spin chain. This 
impurity model is an exactly solved one and it degenerates to the 
integrable $XXZ$ impurity model under the triangular limit. This 
approach 
is the first time to add the integrable impurities to the completely 
anisotropic Heisenberg spin model with the open boundary conditions. 

\end{minipage}
\end{center}


As quantum-mechanical model of magnetism, Heisenberg's 
model\cite{1928} is
very fruitful in the theory of magnetism and is actively studied after
Bethe's work on isotropic case of the $XXX$ model. Yang and 
Yang\cite{yang}
generalized Bethe's method to the $XXZ$ model. Baxter\cite{bbb} in his
remarkable papers gave a solution for the completely anisotropic $XYZ$ 
model
with the use of the Bethe ansatz method. Faddeev and Takhtajan proposed 
the
quantum inverse scattering method for the $XYZ$ spin model and 
simplified
Baxter's formulae\cite{fadde}. Then, many exactly solved models have 
been
proposed and solved by the coordinate Bethe ansatz\cite{hh99}, the
functional Bethe ansatz\cite{hh77,hh88} and algebraic Bethe ansatz 
method 
\cite{hh55,hh66}, etc. Recently the greatest progress has been made 
for the
quantum impurity problems such as Kondo problem and tunneling in 
quantum
wires for the one dimensional electron systems. We know that the 
impurities
play an important role in the strongly correlated electron systems and 
even
a small amount of defects may change the properties of the electron 
systems.
Then it is very important to construct the integrable systems including 
the
impurities. The pioneering work on the impurity model with the 
integrability
was carried out by Andrei and Johannesson\cite{m9} for the isotropic
Heisenberg chain. It were extended to the Babujian-Takhtajan spin chain 
in
Refs. \cite{323301}. Bed\"{u}fig, E$\beta ler$ and Frahm\cite{bef2} 
solved
the integrable model with the impurity coupled with periodic $t-J$ 
chain\cite
{last01,last02,last03}. Schlottmann and Zvyagin have introduced the 
impurity
in supersymmetric $t-J$ model via its scattering matrix with the 
itinerant
electrons\cite{schlo01,schlo02} . The Hamiltonian of the system and 
other
conserved currents can be constructed in principle by the transfer 
matrix.
They have discussed also the magnetic impurities embedded in the 
Hubbard
model\cite{021sch} and a finite concentration of magnetic impurities
embedded in one-dimensional lattice via scattering 
matrices\cite{verry}.

Exactly solved systems with the open boundary conditions have been 
studied
earlier in Refs. [20-24] and a general approach to construct open 
quantum
spin model is given by Cherednik\cite{111} and Sklyanin\cite{222}. 
Then many
exactly solved model with the boundary conditions have been proposed 
and
solved after these pioneering work [27-42]. We notice that the 
impurities
may cut the one-dimensional system into the small part when they are
introduced and (then) the open boundary systems are formed with the
impurities at the ends of the systems. Then the integrable impurity 
model
[43-49] can be constructed from the open boundary system. In this paper, 
we
devote to construct an integrable Hamiltonian of the impurity model 
for the
completely anisotropic $XYZ$ Heisenberg spin chain where the 
impurities are
coupled to the ends of the system. As is well known, the $XYZ\;$ spin 
model
is described by the Hamiltonian 
\begin{equation}
H=\sum_{n=1}^{N-1}\left( J_1\sigma _n^1\sigma _{n+1}^1+J_2\sigma 
_n^2\sigma
_{n+1}^2+J_3\sigma _n^3\sigma _{n+1}^3\right)
\end{equation}
where $J_1,$ $J_2$ and $J_3$ are constants; the spin operators $\sigma 
_n^j$
have the form 
\[
\sigma _n^j=I\otimes \cdots \otimes \sigma ^j\otimes \cdots \otimes 
I\quad
\left( j=1,2,3;\ n=1,\cdots ,N\right) 
\]
where $\sigma ^j$ at site $n$ are the Pauli operators. This Hamiltonian
denotes a one-dimensional quantum mechanical model of the 
ferromagnetism.
There are $N$ spins labelled by $n=1,2,\cdots ,N$ on a line. Every spin 
is
associated with the three-dimensional vector 
$\overrightarrow{\sigma }
=\left( \sigma ^1,\sigma ^2,\sigma ^3\right) $ of the Pauli matrices. 
The
interactions between the neighboring spins are expressed by the 
constants $
J_{1,2,3}.$ The $R$ matrix of the system has the form

\begin{equation}
R\left( \lambda \right) =\left( 
\begin{array}{cccc}
$sn$(\lambda +\eta ) & 0 & 0 & k$sn$\eta $sn$\lambda $sn$(\lambda +\eta ) 
\\ 
0 & $sn$\lambda  & $sn$\eta  & 0 \\ 
0 & $sn$\eta  & $sn$\lambda  & 0 \\ 
k$sn$\eta $sn$\lambda $sn$(\lambda +\eta ) & 0 & 0 & $sn$(\lambda 
+\eta )
\end{array}
\right) ,
\end{equation}
which satisfies the Yang-Baxter equation 
\[
R_{12}\left( \lambda _1-\lambda _2\right) R_{13}\left( \lambda _1-
\lambda
_3\right) R_{23}\left( \lambda _2-\lambda _3\right) \ \qquad \qquad 
\]
\begin{equation}
=R_{23}\left( \lambda _2-\lambda _3\right) R_{13}\left( \lambda 
_1-\lambda
_3\right) R_{12}\left( \lambda _1-\lambda _2\right) 
\end{equation}
and has the properties 
\[
R_{12}\left( \lambda \right) R_{12}\left( -\lambda \right) =\rho 
\left(
\lambda \right) I,
\]
\begin{equation}
R_{12}\left( 0\right) ={\rm sn}\eta P_{12},R_{12}^{t_1t_2}
\left( \lambda \right)
=R_{12}\left( \lambda \right) ,
\end{equation}
\[
R_{12}^{t_1}\left( \lambda \right) R_{12}^{t_1}\left( -\lambda -\eta 
\right)
=\rho \left( \lambda +\eta \right) I
\]
with $\rho \left( \lambda \right) =$sn$^2\eta -$sn$^2\lambda .$ 
Another kind
of the basic properties in the above  Yang-Baxter equation is the 
difference
property that the $R$ matrices rely on only the differences of the
corresponding spectrum parameters, which is useful to construct the 
models
with integrable impurities. The boundary $K$ matrices satisfy the 
reflection
equation\cite{111,222} 
\begin{eqnarray}
&&R_{12}\left( \lambda _1-\lambda _2\right) 
\stackrel{1}{K}_{-}\left(
\lambda _1\right) R_{12}\left( \lambda _1+\lambda _2\right) 
\stackrel{2}{K}
_{-}\left( \lambda _2\right)   \nonumber \\
&=&\stackrel{2}{K}_{-}\left( \lambda _2\right) R_{12}\left( \lambda
_1+\lambda _2\right) \stackrel{1}{K}_{-}\left( \lambda _1\right)
R_{12}\left( \lambda _1-\lambda _2\right)   \label{re001}
\end{eqnarray}
\begin{eqnarray}
&&R_{12}\left( -\lambda _1+\lambda _2\right) 
\stackrel{1}{K}_{+}^{t_1}\left(
\lambda _1\right) R_{12}\left( -\lambda _1-\lambda _2-2\eta \right) 
\stackrel{2}{K}_{+}^{t_2}\left( \lambda _2\right)   \nonumber \\
&=&\stackrel{2}{K}_{+}^{t_2}\left( \lambda _2\right) 
R_{12}\left( -\lambda
_1-\lambda _2-2\eta \right) \stackrel{1}{K}_{+}^{t_1}\left( \lambda
_1\right) R_{12}\left( -\lambda _1+\lambda _2\right)   \label{re002}
\end{eqnarray}
where $\stackrel{1}{K}_{\pm }\equiv K_{\pm }\otimes id_{V_2}$ and 
$\stackrel{
2}{K}_{\pm }\equiv id_{V_1}\otimes $ $K_{\pm }.$ Here I follow the same
notations as in Refs. [26-29]. The transfer matrix $t\left( \lambda 
\right) $
is defined as  
\begin{equation}
t\left( \lambda \right) =tr\left\{ K_{+}\left( \lambda \right) T\left(
\lambda \right) K_{-}\left( \lambda \right) T^{-1}\left( -\lambda 
\right)
\right\} 
\end{equation}
with the use of the boundary $K_{\pm }$ matrices and the monodromy 
matrix $
T\left( \lambda \right) $. It has the commuting property 
\[
\left[ t\left( \lambda \right) ,t\left( \lambda ^{\prime }\right) 
\right] =0,
\]
which can be obtained from the Yang-Baxter equation and the boundary
reflection equations ( \ref{re001}) and (\ref{re002}). This property 
ensures
the integrability of the system with the open boundary conditions. For
convenience, we use the notation 
\begin{equation}
R\left( \lambda \right) =\sum_{j=1}^4W_j\left( \lambda \right) \sigma
^j\otimes \sigma ^j
\end{equation}
where

\begin{eqnarray}
W_4\left( \lambda \right) +W_3
\left( \lambda \right)  &=&{\rm sn}(\lambda +\eta ),
\nonumber \\
W_4\left( \lambda \right) -W_3\left( \lambda \right)  &=&{\rm 
sn}\lambda , 
\nonumber \\
W_1\left( \lambda \right) +W_2\left( \lambda \right)  &=&{\rm sn}\eta , 
\\
W_1\left( \lambda \right) -W_2
\left( \lambda \right)  &=&k{\rm sn}\eta {\rm sn}\lambda
{\rm sn}(\lambda +\eta ).  \nonumber
\end{eqnarray}
In order to add the magnetic impurities to the $XYZ$ spin chain, we 
define
that 
\begin{equation}
T_a\left( \lambda \right) =R_{aR}\left( \lambda +c_R\right) 
R_{aN}\left(
\lambda \right) R_{aN-1}\left( \lambda \right) \cdots 
R_{a1}\left( \lambda
\right) .
\end{equation}
Then the inverse of the monodromy matrix is 
\[
T_a^{-1}\left( -\lambda \right) =\frac 1{\rho ^N\left( \lambda \right) 
\rho
\left( \lambda -c_R\right) }R_{a1}\left( \lambda \right) R_{a2}\left(
\lambda \right) \cdots R_{aN}\left( \lambda \right) 
R_{aR}\left( \lambda
-c_R\right) .
\]
We choice that the boundary $K$ matrices as 
\begin{eqnarray}
K_{-}\left( \lambda \right)  &=&R_{aL}\left( \lambda +c_L\right)
R_{aL}\left( \lambda -c_L\right) ,  \nonumber \\
K_{+}\left( \lambda \right)  &=&1,
\end{eqnarray}
which satisfy the reflection equations. Here the impurity parameters 
$c_R$
and $c_L$ are introduced. They are related to the exchange constants 
between
the particles of the system and the impurities situated at the ends 
of the
chain. These exchange couplings between the boundary spins can be 
turned
away from their bulk values by variation of the parameters $c_R$ and 
$c_L$
respectively. It occurs since the construction is based completely on 
the
difference properties of the solutions of the Yang-Baxter equation (3). 
The
transfer matrix is 
\[
t\left( \lambda \right) =Tr\left\{ T_a\left( \lambda \right) 
K_{-}\left(
\lambda \right) T_a^{-1}\left( -\lambda \right) \right\} .
\]
Then we have that 
\begin{eqnarray}
&&Tr_a\left\{ \left. \frac{dR_{aR}\left( \lambda 
+c_R\right) }{d\lambda }
\right| _{\lambda =0}R_{aN}\left( 0\right) R_{aN-1}\left( 0\right) 
\cdots
R_{a1}\left( 0\right) K_{-}\left( 0\right) T_a^{-1}\left( 0\right) 
\right\} 
\nonumber \\
&=&\frac{2\rho \left( c_L\right) }{\rho \left( c_R\right) }
\sum_{j=1}^4\left. \frac{dW_j\left( \lambda 
+c_R\right) }{d\lambda }\right|
_{\lambda =0}W_j\left( -c_R\right) ,
\end{eqnarray}
\begin{eqnarray}
&&Tr_a\left\{ R_{aR}\left( c_R\right) \left. 
\frac{dR_{aN}\left( \lambda
\right) }{d\lambda }\right| _{\lambda =0}R_{aN-1}\left( 0\right) 
\cdots
R_{a1}\left( 0\right) K_{-}\left( 0\right) T_a^{-1}
\left( 0\right) \right\} 
\nonumber \\
&=&\frac{\rho \left( c_L\right) }{{\rm sn}\eta \rho 
\left( c_R\right) }\left\{
2J_1A_R\sigma _N^1\sigma _R^1+2J_2B_R\sigma _N^2\sigma 
_R^2+2J_3C_R\sigma
_N^3\sigma _R^3\right.   \nonumber \\
&&\left. +2J_3\sum_{j=1}^4W_j\left( c_R\right) W_j\left( -c_R\right)
\right\} ,
\end{eqnarray}
where 
\begin{eqnarray}
A_R &=&W_1\left( c_R\right) W_4\left( -c_R\right) 
+W_4\left( c_R\right)
W_1\left( -c_R\right)  \\
&&+W_2\left( c_R\right) W_3\left( -c_R\right) +W_3\left( c_R\right)
W_2\left( -c_R\right) ,  \nonumber
\end{eqnarray}
\begin{eqnarray}
B_R &=&W_1\left( c_R\right) W_3\left( -c_R\right) 
+W_3\left( c_R\right)
W_1\left( -c_R\right)   \nonumber \\
&&+W_2\left( c_R\right) W_4\left( -c_R\right) +W_4\left( c_R\right)
W_2\left( -c_R\right) ,
\end{eqnarray}
\begin{eqnarray}
C_R &=&W_1\left( c_R\right) W_2\left( -c_R\right) 
+W_2\left( c_R\right)
W_1\left( -c_R\right)   \nonumber \\
&&+W_3\left( c_R\right) W_4\left( -c_R\right) +W_4\left( c_R\right)
W_3\left( -c_R\right) .
\end{eqnarray}
By the use of 
\[
P_{nn+1}\left. \frac{dR_{nn+1}\left( \lambda 
\right) }{d\lambda }\right|
_{\lambda =0}=\frac{1+k{\rm sn}^2\eta }2\sigma _n^1\sigma 
_{n+1}^1+\frac{
1-k{\rm sn}^2\eta }2\sigma _n^2\sigma _{n+1}^2
\]
\begin{equation}
+\frac{{\rm cn}\eta {\rm dn}\eta }2\sigma _n^3\sigma _{n+1}^3+
\frac{{\rm cn}\eta {\rm dn}\eta }2,
\end{equation}
we have that 
\begin{eqnarray}
&&\sum_{j=1}^{N-1}Tr_a\left\{ R_{aR}\left( c_R\right) 
R_{aN}\left( 0\right)
\cdots R_{aj+1}\left( 0\right) \left. \frac{dR_{aj}
\left( \lambda \right) }{
d\lambda }\right| _{\lambda =0}\right.   \nonumber \\
&&\cdot \left. R_{aj-1}\left( 0\right) \cdots R_{a1}\left( 0\right)
K_{-}\left( 0\right) T_a^{-1}\left( 0\right) \right\}   \nonumber \\
&=&\frac{2\rho \left( c_L\right) }{{\rm sn}\eta }
\sum_{j=1}^{N-1}\left( J_1\sigma
_n^1\sigma _{n+1}^1+J_2\sigma _n^2\sigma _{n+1}^2+J_3\sigma 
_n^3\sigma
_{n+1}^3+J_3\right) 
\end{eqnarray}
where $J_1=\left( 1+k{\rm sn}^2\eta \right) /2,$ $J_2=
\left( 1-k{\rm sn}^2\eta \right) /2
$ and $J_3=$cn$\eta $dn$\eta /2$ ; and 
\begin{eqnarray}
&&\frac 1{\rho ^N\left( \lambda \right) \rho \left( \lambda -
c_R\right)
}Tr_a\left\{ T_a\left( 0\right) K_{-}
\left( 0\right) R_{a1}\left( 0\right)
\cdots \right.   \nonumber \\
&&\cdot \left. R_{aN}\left( 0\right) \left. 
\frac{dR_{aR}\left( \lambda
+c_R\right) }{d\lambda }\right| _{\lambda =0}\right\}   \nonumber \\
&=&\frac{2\rho \left( c_L\right) }{\rho \left( c_R\right) }
\sum_{j=1}^4W_j\left( c_R\right) \left. \frac{dW_j
\left( \lambda -c_R\right) 
}{d\lambda }\right| _{\lambda =0},
\end{eqnarray}
\begin{eqnarray}
&&Tr_a\left\{ T_a\left( 0\right) \left. \frac{dK_{-}
\left( \lambda \right) }{
d\lambda }\right| _{\lambda =0}T_a^{-1}
\left( 0\right) \right\}   \nonumber
\\
&=&2A_L\sigma _1^1\sigma _L^1+2B_L\sigma _1^2\sigma _L^2+2C_L\sigma
_1^3\sigma _L^3+D_L
\end{eqnarray}
where 
\begin{eqnarray}
A_L &=&\left. \frac{dW_1\left( \lambda 
+c_L\right) }{d\lambda }\right|
_{\lambda =0}W_4\left( -c_L\right) +\left. \frac{dW_4\left( \lambda
+c_L\right) }{d\lambda }\right| _{\lambda =0}W_1\left( -c_L\right)  
\nonumber \\
&&-\left. \frac{dW_2\left( \lambda +c_L\right) }
{d\lambda }\right| _{\lambda
=0}W_3\left( -c_L\right) -\left. \frac{dW_3
\left( \lambda +c_L\right) }{
d\lambda }\right| _{\lambda =0}W_2\left( -c_L\right)   \nonumber \\
&&+\left. \frac{dW_1\left( \lambda -c_L\right) }
{d\lambda }\right| _{\lambda
=0}W_4\left( c_L\right) +\left. 
\frac{dW_4\left( \lambda -c_L\right) }{
d\lambda }\right| _{\lambda =0}W_1\left( c_L\right)   \nonumber \\
&&-\left. \frac{dW_2\left( \lambda -c_L\right) }
{d\lambda }\right| _{\lambda
=0}W_3\left( c_L\right) -\left. \frac{dW_3
\left( \lambda -c_L\right) }{
d\lambda }\right| _{\lambda =0}W_2\left( c_L\right) ,
\end{eqnarray}
\begin{eqnarray}
B_L &=&\left. \frac{dW_2\left( \lambda 
+c_L\right) }{d\lambda }\right|
_{\lambda =0}W_4\left( -c_L\right) +\left. \frac{dW_4\left( \lambda
+c_L\right) }{d\lambda }\right| _{\lambda =0}W_2\left( -c_L\right)  
\nonumber \\
&&-\left. \frac{dW_1\left( \lambda +c_L\right) }
{d\lambda }\right| _{\lambda
=0}W_3\left( -c_L\right) -\left. 
\frac{dW_3\left( \lambda +c_L\right) }{
d\lambda }\right| _{\lambda =0}W_1
\left( -c_L\right)   \nonumber \\
&&+\left. \frac{dW_2\left( \lambda -c_L\right) }
{d\lambda }\right| _{\lambda
=0}W_4\left( c_L\right) +\left. 
\frac{dW_4\left( \lambda -c_L\right) }{
d\lambda }\right| _{\lambda =0}W_2\left( c_L\right)   \nonumber \\
&&-\left. \frac{dW_1\left( \lambda -c_L\right) }
{d\lambda }\right| _{\lambda
=0}W_3\left( c_L\right) -\left. 
\frac{dW_3\left( \lambda -c_L\right) }{
d\lambda }\right| _{\lambda =0}W_1\left( c_L\right) ,
\end{eqnarray}
\begin{eqnarray}
C_L &=&\left. \frac{dW_3
\left( \lambda +c_L\right) }{d\lambda }\right|
_{\lambda =0}W_4\left( -c_L\right) +
\left. \frac{dW_4\left( \lambda
+c_L\right) }{d\lambda }\right| _{\lambda =0}W_3\left( -c_L\right)  
\nonumber \\
&&-\left. \frac{dW_1\left( \lambda +c_L\right) }
{d\lambda }\right| _{\lambda
=0}W_2\left( -c_L\right) -\left. 
\frac{dW_2\left( \lambda +c_L\right) }{
d\lambda }\right| _{\lambda =0}W_1
\left( -c_L\right)   \nonumber \\
&&+\left. \frac{dW_3\left( \lambda -c_L\right) }
{d\lambda }\right| _{\lambda
=0}W_4\left( c_L\right) +\left. \frac{dW_4\left( \lambda -
c_L\right) }{
d\lambda }\right| _{\lambda =0}W_3\left( c_L\right)   \nonumber \\
&&-\left. \frac{dW_1\left( \lambda -c_L\right) }
{d\lambda }\right| _{\lambda
=0}W_2\left( c_L\right) -\left. \frac{dW_2\left( \lambda -
c_L\right) }{
d\lambda }\right| _{\lambda =0}W_1\left( c_L\right) ,
\end{eqnarray}
\begin{equation}
D_L=\sum_{j=1}^4\left\{ \left. \frac{dW_j
\left( \lambda +c_L\right) }{
d\lambda }\right| _{\lambda =0}W_j\left( -c_L\right) 
+W_j\left( c_L\right)
\left. \frac{dW_j\left( \lambda -c_L\right) }
{d\lambda }\right| _{\lambda
=0}\right\} .
\end{equation}
Therefore we get that 
\begin{eqnarray}
\frac{{\rm sn}\eta }{4\rho \left( c_L\right) }
\left. \frac{dt\left( \lambda
\right) }{d\lambda }
\right| _{\lambda =0} &=&\sum_{n=1}^{N-1}\left(
J_1\sigma _n^1\sigma _{n+1}^1+
J_2\sigma _n^2\sigma _{n+1}^2+J_3\sigma
_n^3\sigma _{n+1}^3\right)   \nonumber \\
&&+\frac{{\rm sn}\eta }{2\rho \left( c_L\right) }
\left( A_L\sigma _1^1\sigma
_L^1+B_L\sigma _1^2\sigma _L^2
+C_L\sigma _1^3\sigma _L^3\right)   \nonumber
\\
&&+\frac 1{\rho \left( c_R\right) }\left( J_1A_R\sigma _N^1\sigma
_R^1+J_2B_R\sigma _N^2\sigma _R^2+J_3C_R\sigma _N^3\sigma 
_R^3\right)  
\nonumber \\
&&+\frac{{\rm sn}\eta }{2\rho \left( c_L\right) }C_L+
\frac{{\rm sn}\eta }{2\rho \left(
c_R\right) }C_L+\frac{J_3}{\rho \left( c_R\right) }
\sum_{j=1}^4W_j\left(
c_R\right) W_j\left( -c_R\right)   \nonumber \\
&&+\left( N-1\right) J_3.
\end{eqnarray}
Set 
\begin{equation}
H\equiv \frac{{\rm sn}\eta }{4{\rm sn}^2\eta -4{\rm sn}^2c_L}
\left. \frac{dt\left( \lambda
\right) }{d\lambda }\right| _{\lambda =0}-h_0
\end{equation}
with 
\begin{eqnarray}
h_0 &=&NJ_3+\frac 14{\rm sn}\eta {\rm sn}\left( 2\eta \right) 
\left( 1-k^2{\rm sn}^2\eta
{\rm sn}^2c_L\right)   \nonumber \\
&&\cdot \left[ 1-k^2{\rm sn}^2\left( \eta +c_L\right) {\rm sn}^2
\left( \eta -c_L\right)
\right]   \nonumber \\
&&\cdot \left( \frac 1{{\rm sn}^2\eta -{\rm sn}^2c_L}+
\frac 1{{\rm sn}^2\eta -{\rm sn}^2c_R}\right) .
\end{eqnarray}
We get the following Hamiltonian of the impurity model, 
\begin{eqnarray}
H &=&\sum_{n=1}^{N-1}\left( J_1\sigma _n^1\sigma _{n+1}^1+J_2\sigma
_n^2\sigma _{n+1}^2+J_3\sigma _n^3\sigma _{n+1}^3\right)   \nonumber 
\\
&&+\frac 1{{\rm sn}^2\eta -{\rm sn}^2c_L}\left\{ J_{L,1}\sigma 
_1^1\sigma
_L^1+J_{L,2}\sigma _1^2\sigma _L^2+J_{L,3}\sigma _1^3\sigma 
_L^3\right\}  
\nonumber \\
&&+\frac 1{{\rm sn}^2\eta -{\rm sn}^2c_R}\left\{ J_{R,1}\sigma 
_N^1\sigma
_R^1+J_{R,2}\sigma _N^2\sigma _R^2+J_{R,3}\sigma _N^3\sigma 
_R^3\right\} ,
\end{eqnarray}
where 
\[
J_1=\frac{1+k{\rm sn}^2\eta }2,\quad J_2=
\frac{1-k{\rm sn}^2\eta }2,\quad J_3=\frac{
{\rm cn}\eta {\rm dn}\eta }2,
\]
\[
J_{L,1}=\frac 12{\rm sn}^2\eta {\rm cn}c_L{\rm dn}c_L
\left[ 1+k{\rm sn}\left( \eta +c_L\right)
{\rm sn}\left( \eta -c_L\right) \right] ,
\]
\[
J_{L,2}=\frac 12{\rm sn}^2\eta {\rm cn}c_L{\rm dn}c_L
\left[ 1-k{\rm sn}\left( \eta +c_L\right)
{\rm sn}\left( \eta -c_L\right) \right] ,
\]
\begin{eqnarray*}
J_{L,3} &=&\frac 14{\rm sn}\eta {\rm sn}\left( 2\eta \right) 
\left( 1-k^2{\rm sn}^2\eta
{\rm sn}^2c_L\right)  \\
&&\cdot \left[ 1-k^2{\rm sn}^2\left( \eta +c_L\right) {\rm sn}^2
\left( \eta -c_L\right)
\right] .
\end{eqnarray*}
And the exchange constants $J_{R,i}$ have the same expressions as the
constants $J_{L,i}$ ( $i=1,2,3$ ) except for the substitution of the
parameters $c_L$ by $c_R.$ It is an integrable impurity model. The two
impurities are coupled to both of the ends of the completely anisotropic
Heisenberg spin chain. When we take the triangular limit $k\rightarrow 
0$ (
the supplementary modulus $k^{\prime }$ tends to $1$ ), the elliptic
functions sn$u,$cn$u,$dn$u$ become $\sin u,$ $\cos u,$ and $1,$
respectively. The above Hamiltonian reduces to 
\begin{eqnarray}
H &=&\frac 12\left\{ \sum_{n=1}^{N-1}\left( \sigma _n^1\sigma
_{n+1}^1+\sigma _n^2\sigma _{n+1}^2+\cos \eta \sigma _n^3\sigma
_{n+1}^3\right) \right.   \nonumber \\
&&+\frac{\sin ^2\eta \cos c_L}{\sin ^2\eta -\sin ^2c_L}\left( \sigma
_1^1\sigma _L^1+\sigma _1^2\sigma _L^2+\frac{\cos \eta }{\cos 
c_L}\sigma
_1^3\sigma _L^3\right)   \nonumber \\
&&+\frac{\sin ^2\eta \cos c_R}{\sin ^2\eta -\sin ^2c_R}\left( \sigma
_N^1\sigma _R^1+\sigma _N^2\sigma _R^2+\frac{\cos \eta }{\cos 
c_R}\sigma
_N^3\sigma _R^3\right) .
\end{eqnarray}
It is just the Hamiltonian of the impurity model related to $XXZ\;$ 
spin
model\cite{many09,chen}.

As the conclusion, we have added the magnetic impurities to the edges 
of the
completely anisotropic Heisenberg $XYZ$ spin model. The model is 
exactly
solvable and the Hamiltonian of the impurity model is obtained 
explicitly.
The interactions between the impurities and the electrons are described 
by
the arbitrary parameters $c_L$ and $c_R,$ which comes to the difference
properties of the spectrum parameters of the $R$ matrices of the spin 
chain.
Under the triangular limit, our Hamiltonian of the impurity model 
reduces to
the one related to the $XXZ$ spin model\cite{many09}. We known that 
the $XYZ$
spin model is equivalent to the eight vertex model\cite{hh77} and some
progress has been made for the eight vertex models with the open 
boundary
conditions\cite{fan,jimbo}. Batchelor $et$ $al$\cite{hh17014} have 
obtained
the surface free energy with the boundary $K$ matrix\cite{h34}. It is 
an
interesting subject to discuss the impurity effects in the two 
dimensional
lattice model in statistical mechanics. Furthermore, the Bethe ansatz
equations of this system with the impurities can be obtained by the 
use of
the general Bethe ansatz method. We may study the thermodynamics of 
the $XYZ$
model including the magnetic impurities and the impurity effects in 
the
ground state and excited state. All of these are in investigations.

\textheight=240mm

\end{document}